\documentclass[onecolumn,aps,prd,preprintnumbers,showpacs,superscriptaddress,nofootinbib,amsmath,amssymb,floats,floatfix,showkeys,notitlepage,longbi]{revtex4}
\usepackage{mwe,tikz}
\usepackage{orcidlink}
\usepackage{lipsum}
\usepackage{graphicx}
\usepackage{subfigure}
\usepackage{sans}
\usepackage{changes}
\usepackage{hyperref}
\hypersetup{colorlinks=true,linkcolor=blue,urlcolor=blue,citecolor=blue}
\usepackage[toc,page]{appendix}
\usepackage[normalem]{ulem}
\usepackage{adjustbox}
\usepackage{latexsym}
\usepackage{amsmath}
\usepackage{amssymb}
\usepackage{amsfonts}
\usepackage{dcolumn}
\usepackage{bm}
\usepackage{tikz}
\usepackage{bigints}
\usepackage{array,tabularx,multirow}
\usepackage[tracking=true]{microtype}
\usepackage{mathrsfs}
\usepackage{mathtools}
\usepackage{braket}
\usepackage[utf8]{inputenc}
\SetTracking{}{500}
\SetTracking{encoding={*}, shape=sc}{40}
\UseRawInputEncoding %for inputenc error%
\allowdisplaybreaks

\newcommand{\be}{\begin{eqnarray}}
\newcommand{\ee}{\end{eqnarray}}

\def\be{\begin{equation}}
\def\ee{\end{equation}}
\def\bestar{\begin{equation*}}
\def\eestar{\end{equation*}}
\newcommand{\bea}{\begin{eqnarray}}\newcommand{\eea}{\end{eqnarray}}
\newcommand{\brr}{\begin{array}}\newcommand{\err}{\end{array}}
\newcommand{\bit}{\begin{itemize}}\newcommand{\eit}{\end{itemize}}
\newcommand{\ben}{\begin{enumerate}}\newcommand{\een}{\end{enumerate}}

\newcommand{\ba}{\begin{array}}
\newcommand{\ea}{\end{array}}

\graphicspath{{./images/} }

\newcolumntype{M}[1]{>{\centering\arraybackslash}m{#1}}
\newcolumntype{N}{@{}m{0pt}@{}}

\newcounter{sxn}

\newcounter{axn}

\newdimen\mybaselineskip
\newcommand{\beeq}{\begin{equation}}
\newcommand{\eneq}{\end{equation}}
\newcommand{\beqn}{\begin{eqnarray}}
\newcommand{\eeqn}{\end{eqnarray}}

\newcommand{\beal}{\setcounter{letter}{1} \begin{eqnarray}}
\newcommand{\eeal}{\addtocounter{equation}{1} \end{eqnarray}}

\newcommand{\larrow}{\,\,\,\,\hbox to 30pt{\rightarrowfill}
\,\,\,\,}
\newcommand{\slarrow}{\,\,\,\hbox to 20pt{\rightarrowfill}
\,\,\,}

\def\la{\raise.16ex\hbox{$\langle$}\lower.16ex\hbox{}  }
\def\ra{\, \raise.16ex\hbox{$\rangle$}\lower.16ex\hbox{} }

\def\psibar{ \psi \kern-.65em\raise.6em\hbox{$-$} \lower.6em\hbox{} }
\def\psibarb{ \psi \kern-.65em\raise.6em\hbox{$-$}  }

\begin{document} 

\title{Probing Schwarzschild-like Black Holes in Metric-Affine Bumblebee Gravity with Accretion Disk, Deflection Angle, Greybody Bounds, and Neutrino Propagation
}

%order as a surnames

\author{Gaetano Lambiase}
\email{lambiase@sa.infn.it}
\affiliation{Dipartimento di Fisica ``E.R Caianiello'', Università degli Studi di Salerno, Via Giovanni Paolo II, 132 - 84084 Fisciano (SA), Italy.}
\affiliation{Istituto Nazionale di Fisica Nucleare - Gruppo Collegato di Salerno - Sezione di Napoli, Via Giovanni Paolo II, 132 - 84084 Fisciano (SA), Italy.}

\author{Leonardo Mastrototaro}
\email{lmastrototaro@unisa.it}
\affiliation{Dipartimento di Fisica ``E.R Caianiello'', Università degli Studi di Salerno, Via Giovanni Paolo II, 132 - 84084 Fisciano (SA), Italy.}
\affiliation{Istituto Nazionale di Fisica Nucleare - Gruppo Collegato di Salerno - Sezione di Napoli, Via Giovanni Paolo II, 132 - 84084 Fisciano (SA), Italy.}

\author{Reggie C. Pantig}
%\orcidlink{0000- 0002-3101-8591}
\email{rcpantig@mapua.edu.ph}
\affiliation{Physics Department, Map\'ua University, 658 Muralla St., Intramuros, Manila 1002, Philippines}

\author{Ali \"Ovg\"un}
%\orcidlink{0000- 0002-9889-342X}
\email{ali.ovgun@emu.edu.tr}
\affiliation{Physics Department, Eastern Mediterranean University, Famagusta, 99628 North
Cyprus via Mersin 10, Turkey.}

\begin{abstract}
In this paper, we investigate Schwarzschild-like black holes within the framework of metric-affine bumblebee gravity. We explore the implications of such a gravitational setup on various astrophysical phenomena, including the presence of an accretion disk, the deflection angle of light rays, the establishment of greybody bounds, and the propagation of neutrinos. The metric-affine bumblebee gravity theory offers a unique perspective on gravitational interactions by introducing a vector field that couples to spacetime curvature. We analyze the behavior of accretion disks around Schwarzschild-like black holes in this modified gravity scenario, considering the effects of the bumblebee field on the accretion process. Furthermore, we scrutinize the deflection angle of light rays as they traverse the gravitational field, highlighting potential deviations from standard predictions due to the underlying metric-affine structure. Investigating greybody bounds in this context sheds light on the thermal radiation emitted by black holes and how the modified gravity framework influences this phenomenon. Moreover, we explore neutrino propagation around Schwarzschild-like black holes within metric-affine bumblebee gravity, examining alterations in neutrino trajectories and interactions compared to conventional general relativity. By comprehensively probing these aspects, we aim to unravel the distinctive features and consequences of Schwarzschild-like black holes in the context of metric-affine bumblebee gravity, offering new insights into the nature of gravitational interactions and their observable signatures.
\end{abstract}

\date{\today}


\maketitle
%\tableofcontents

\section{Introduction} \label{sec1}

A significant hurdle in the field of theoretical physics involves the harmonization of Einstein's widely accepted theory of gravitation, known as general relativity (GR), with the standard model of particle physics (SM), which adeptly brings together all other fundamental forces. One potential avenue for addressing this quandary lies in the concept of spontaneous symmetry disruption, a pivotal factor in the realm of elementary particle physics. During the initial stages of the universe, it's plausible that the temperature reached a point where such symmetry disruption could have been activated.

One form of symmetry disruption that arises while attempting to quantize general relativity is the breakdown of Lorentz symmetry. Studies reveal that this symmetry might experience significant violations at the Planck scale (around $10^{19} GeV$) in various approaches to quantum gravity (QG), indicating its potential non-fundamental nature in the natural order. Additionally, this Lorentz symmetry breakdown (LSB) could furnish conceivable indicators of the underlying quantum gravity framework at lower energy levels. However, implementing consistent Lorentz symmetry breakdown (LSB) within the gravitational context presents distinct challenges when compared to incorporating Lorentz-breaking extensions into non-gravitational field theories. In the context of flat spacetimes, it's feasible to introduce additive terms that break Lorentz symmetry, such as the Carroll-Field-Jackiw term \cite{Carroll:1989vb}, aether time \cite{Carroll:2008pk}, and other analogous terms (as seen in \cite{Colladay:1998fq}). These terms can be rooted in a constant vector (tensor) multiplied by functions of fields and their derivatives. Nevertheless, when dealing with curved spacetimes, these features cannot be suitably adapted. Nevertheless, effects stemming from the underlying quantum gravity theory may manifest themselves at lower energy scales, offering glimpses of this grand unification.
In 1989, Kostelecky and Samuel pioneered a simple model for spontaneous Lorentz violation known as bumblebee gravity \cite{Kostelecky:1989jw}. In this model, a bumblebee field with a vacuum expectation value disrupts Lorentz symmetry, and Lorentz violation emerges from the dynamics of a single vector field, denoted as $B_\mu$ \cite{Kostelecky:2000mm,Kostelecky:2002ca,Bertolami:2003qs,Cambiaso:2012vb}. Hence one avenue to explore the Planck-scale signals and the potential breaking of relativity is through the violation of Lorentz symmetry \cite{Kostelecky:2020hbb}. Theories that violate Lorentz symmetry at the Planck scale, while incorporating elements of both GR and the SM, are encompassed by effective field theories known as the Standard Model Extension (SME) \cite{Kostelecky:2003fs,Bertolami:2005bh,Casana:2017jkc,Santos:2014nxm,Jesus:2020lsv,Jesus:2019nwi,Maluf:2021lwh,Maluf:2020kgf,KumarJha:2020ivj,Xu:2022frb,Ovgun:2018ran,Li:2020wvn,Yang:2018zef,Ovgun:2018xys,Oliveira:2018oha,Mangut:2023oxa,Gullu:2020qzu,Kuang:2022xjp,Nascimento:2023auz,Delhom:2022xfo,Delhom:2020gfv,Brito:2020eiy,Assuncao:2019azw,Assuncao:2017tnz,Marques:2023suh,Khodadi:2022pqh}.

Einstein's theory of general relativity has proven its mettle through numerous experimental tests, including the groundbreaking technique of gravitational lensing \cite{Cunha:2018acu}. Gravitational lensing not only helps us comprehend galaxies, dark matter, dark energy, and the universe but also plays a crucial role in understanding black holes, wormholes, global monopoles, and other celestial objects \cite{Virbhadra:1999nm,Virbhadra:2002ju,Virbhadra:2007kw,Virbhadra:2008ws,Keeton:1997by,Bozza:2002zj,Eiroa:2002mk,Sharif:2015qfa,Pantig:2022toh,Pantig:2022gih,Pulice:2023dqw,Lambiase:2023hng,Ovgun:2023ego,Pantig:2022qak,Rayimbaev:2022hca,Uniyal:2022vdu,Vagnozzi:2022moj,Pantig:2022whj,Okyay:2021nnh,Khodadi:2023yiw,Khodadi:2021owg}. A novel approach to calculate the deflection angle of light has been introduced by Gibbons and Werner. This method allows for the computation of light deflection in non-rotating asymptotically flat spacetimes. It leverages the Gauss-Bonnet theorem within the context of the optical geometry surrounding a black hole \cite{Gibbons:2008rj}. Furthermore, this pioneering technique has been extended to encompass stationary spacetimes by Werner \cite{Werner:2012rc}. Einstein's theory of gravity has yielded one of its most profound predictions: the existence of black holes. The existence of black holes has been firmly established through a wealth of astrophysical observations. Notably, the detection of gravitational waves (GWs) by LIGO/VIRGO collaborations has provided compelling evidence for black holes \cite{LIGOScientific:2016aoc}. Additionally, the Event Horizon Telescope (EHT) has made history by capturing the first images of the shadow cast by supermassive black holes at the centers of galaxies, including M87* and Sgr A* \cite{EventHorizonTelescope:2019dse}. These remarkable achievements have not only confirmed the existence of black holes but have also opened new avenues for the study of their properties and the nature of gravity in the strong-field regime.

Main aim of this paper is to examine the characteristics of Schwarzschild-like black holes within the framework of metric-affine bumblebee gravity \cite{Filho:2022yrk}. We seek to unravel the far-reaching consequences of this gravitational framework on a wide range of astrophysical phenomena. Specifically, we investigate its impact on the formation and behavior of accretion disks, the deflection patterns of light rays, the establishment of greybody bounds, and the propagation behaviors of neutrinos. Through these investigations, we aim to deepen our understanding of the unique properties and effects associated with black holes in metric-affine bumblebee gravity.

The organization of this manuscript is outlined as follows. In Section II, we investigate the weak deflection angle of spherically symmetric metric of Black holes in a metric-affine bumblebee gravity. In the subsequent Section III, we undertake an analysis of the spherically infalling accretion disk exhibited by these spherical black holes in a metric-affine bumblebee gravity  featuring a LSV parameters. In Section IV, we study the greybody factors of the black hole, and investigate the neutrino energy deposition in the Section V. A summary of our investigation is presented in Section VI.

\section{Weak deflection angle of black holes in a metric-affine bumblebee gravity}

In this section, we study the weak deflection angle of spherically symmetric metric of black holes in a metric-affine bumblebee gravity \cite{Filho:2022yrk}

\begin{equation}
\label{eq.metric}
\begin{array}{l}\qquad d s_{(g)}^{2}=-\frac{\left(1-\frac{2 M}{r}\right)}{\sqrt{\left(1+\frac{3 X}{4}\right)\left(1-\frac{X}{4}\right)}} d t^{2}+\frac{d r^{2}}{\left(1-\frac{2 M}{r}\right)} \sqrt{\frac{\left(1+\frac{3 X}{4}\right)}{\left(1-\frac{X}{4}\right)^{3}}}+r^{2}\left(d \theta^{2}+\sin ^{2} \theta d \phi^{2}\right).  \end{array}
\end{equation}

In our analysis, we've employed a condensed symbol, denoted as $X$, which succinctly signifies the Lorentz-violating parameter, and this symbol is defined as $X=\xi b^{2}$.

To compute the weak deflection angle by GBT theorem, it was demonstrated in a non-asymptotic black hole spacetime by Li et al.  \cite{Li:2020wvn} that the GBT can be expressed as follows:
\begin{equation} \label{eIshi}
    \hat{\alpha} = \iint_{D}KdS + \phi_{\text{RS}},
\end{equation}
Here, let's define some key terms: $r_\text{ps}$ represents the radius of the particle's circular orbit, while S and R indicate the radial positions of the source and receiver, respectively. These positions are the integration domains. It is important to note that the infinitesimal curved surface element $dS$ can be expressed as:

\begin{equation}
dS = \sqrt{g} \, dr \, d\phi.
\end{equation}

Additionally, $\phi_\text{RS}$ denotes the coordinate position angle between the source and the receiver, defined as $\phi_\text{RS} = \phi_\text{R} - \phi_\text{S}$. This angle can be determined through an iterative solution of the equation:
\begin{align}
    F(u) = \left(\frac{du}{d\phi}\right)^2  %\nonumber\\
    = \frac{C(u)^2u^4}{A(u)B(u)}\Bigg[\left(\frac{E}{J}\right)^2-A(u)\left(\frac{1}{J^2}+\frac{1}{C(u)}\right)\Bigg].
\end{align}
Then we apply the substitution $r = 1/u$ and derive the angular momentum and energy of the massive particle based on the given impact parameter $b$.
\begin{equation}
    J = \frac{\mu v b}{\sqrt{1-v^2}}, \quad E = \frac{\mu}{\sqrt{1-v^2}}.
\end{equation}
With Eq. \eqref{eq.metric}, we find
\begin{align}
    F(u) = \frac{1}{b^2}-u^2-\frac{\left[1+\left(3 b^{2} u^{2}-3\right) v^{2}\right] X}{4 v^{2} b^{2}} %\nonumber\\
    -\frac{3 \left[1+\left(b^{2} u^{2}-1\right) v^{2}\right] M X u}{2 v^{2} b^{2}} + \frac{2 \left[1+\left(b^{2} u^{2}-1\right) v^{2}\right] M u}{v^{2} b^{2}}.
\end{align}
The above enables one to solve for the azimuthal separation angle $\phi$ as
\begin{align} \label{ephi}
    \phi = \arcsin(bu)+\frac{M\left[v^{2}\left(b^{2}u^{2}-1\right)-1\right]}{bv^{2}\sqrt{1-b^{2}u^{2}}} - \frac{X}{8 v^{2} \sqrt{-b^{2} u^{2}+1}} %\nonumber \\
    + \frac{\left[b^{3} u^{3} v^{2}+2 u^{2} v^{2} b^{2}+\left(-v^{2}+1\right) u b -2 v^{2}\right] X M}{8 \left(-b^{2} u^{2}+1\right)^{\frac{3}{2}} b \,v^{4}},
\end{align}
which is the also the direct expression for $\phi_S$ as $u$ is replaced by $u_S$. Meanwhile, the expression for the receiver is $\phi_R = \pi - \phi_S$ where $u_S$ should be replaced by $u_R$.

Leaving the angle $\phi$ for a while, the Gaussian curvature $K$ in terms of connection coefficients can be calculated as
\begin{align}
    K=\frac{1}{\sqrt{g}}\left[\frac{\partial}{\partial\phi}\left(\frac{\sqrt{g}}{g_{rr}}\Gamma_{rr}^{\phi}\right)-\frac{\partial}{\partial r}\left(\frac{\sqrt{g}}{g_{rr}}\Gamma_{r\phi}^{\phi}\right)\right] %\nonumber \\
    =-\frac{1}{\sqrt{g}}\left[\frac{\partial}{\partial r}\left(\frac{\sqrt{g}}{g_{rr}}\Gamma_{r\phi}^{\phi}\right)\right]
\end{align}
since $\Gamma_{rr}^{\phi} = 0$. If there exists an analytical solution for $r_\text{ps}$ within a particular spacetime, then we can establish the following relationship:
\begin{equation} \label{gct}
    \int_{r_\text{ps}}^{r(\phi)} K\sqrt{g}dr = -\frac{A(r)\left(E^{2}-A(r)\right)C'-E^{2}C(r)A(r)'}{2A(r)\left(E^{2}-A(r)\right)\sqrt{B(r)C(r)}}\bigg|_{r = r(\phi)}
\end{equation}
then
\begin{equation}
    \left[\int K\sqrt{g}dr\right]\bigg|_{r=r_\text{ps}} = 0.
\end{equation}
The prime notation signifies differentiation with respect to the radial coordinate, $r$. Consequently, the weak deflection angle \cite{Li:2020wvn}, is given by:
\begin{align} \label{ewda}
    \hat{\alpha} = \int^{\phi_\text{R}}_{\phi_\text{S}} \left[-\frac{A(r)\left(E^{2}-A(r)\right)C'-E^{2}C(r)A(r)'}{2A(r)\left(E^{2}-A(r)\right)\sqrt{B(r)C(r)}}\bigg|_{r = r(\phi)}\right] %\times \nonumber\\
    d\phi + \phi_\text{RS}.
\end{align}
Then we find
\begin{align} \label{gct2}
    &\left[\int K\sqrt{g}dr\right]\bigg|_{r=r_\phi} = -\phi_\text{RS} -\frac{(\cos \phi_R - \cos \phi_S) \left(v^{2}+1\right) M}{v^{2} b}+\frac{3 X \phi_\text{RS}}{8} \nonumber\\
    &+\frac{X M}{8bv^4} \left[\phi_\text{RS}(1+v^2) + (\cos \phi_R - \cos \phi_S) (v^2 + 3v^4 + 2) \right].
\end{align}
With the above expression, Eq. \eqref{ephi} is needed. One should note that if $\phi_S$ is given, $\phi_\text{RS} = \pi - 2\phi_S$. The cosine of $\phi$ is then
\begin{align} \label{cs}
    \cos\phi = \sqrt{1-b^{2}u^{2}}-\frac{Mu\left[v^{2}\left(b^{2}u^{2}-1\right)-1\right]}{\sqrt{v^{2}\left(1-b^{2}u^{2}\right)}} %\nonumber\\
    + \frac{b u X}{8 v^{2} \sqrt{-b^{2} u^{2}+1}} - \frac{\left[1+\left(2 b^{4} u^{4}+2 b^{3} u^{3}-3 b^{2} u^{2}-2 b u +1\right) v^{2}\right] X M}{8 \left(-b^{2} u^{2}+1\right)^{\frac{3}{2}} v^{4} b}
\end{align}
which should be applied to the source and the receiver. Using the above expression to Eq. \eqref{gct2}, we get the final analytic expression for the weak deflection angle that accommodates both time-like particles and finite distance as
\begin{align} \label{wda1}
    \alpha &= \frac{2M\left(v^{2}+1\right)}{bv^{2}}\left(\sqrt{1-b^{2}u_\text{S}^{2}}+\sqrt{1-b^{2}u_\text{S}^{2}}\right) + \frac{3}{8}X\left[\pi-2(\sin^{-1}(bu_\text{S})+\sin^{-1}(bu_\text{R}))\right] \nonumber \\
    &+\frac{MX}{8bv^{4}}\left\{ \left(v^{2}+1\right)\left[\pi-2(\sin^{-1}(bu_\text{S})+\sin^{-1}(bu_\text{S}))\right]-2\left(3v^{4}+v^{2}+2\right)\left(\sqrt{1-b^{2}u_\text{S}^{2}}+\sqrt{1-b^{2}u_\text{R}^{2}}\right)\right\}.
\end{align}
Assuming that $u_R = u_S$, and these are distant from the black hole ($u \rightarrow \infty$),
\begin{equation} \label{wda2}
    \alpha = \frac{2 \left(v^{2}+1\right) M}{v^{2} b}+\frac{3 X \pi}{8}+\frac{ X M}{8 v^{4} b} \left[-6 v^{4}+\left(\pi -2\right) v^{2}+\pi -4\right].
\end{equation}
Finally, when $v = 1$,
\begin{equation} \label{wda3}
    \alpha = \frac{4 M}{b}+\frac{3 X \pi}{8}+\frac{\left(\pi-6 \right) X M}{4 b}.
\end{equation}

\begin{figure*}
    \centering
    \includegraphics[width=0.48\textwidth]{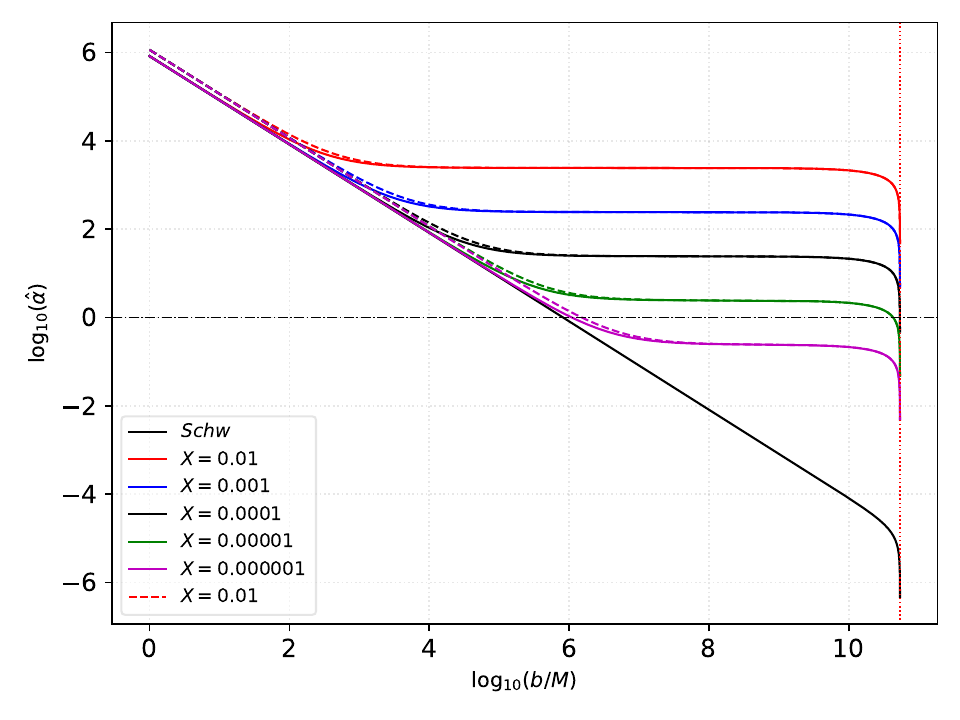}
    \includegraphics[width=0.48\textwidth]{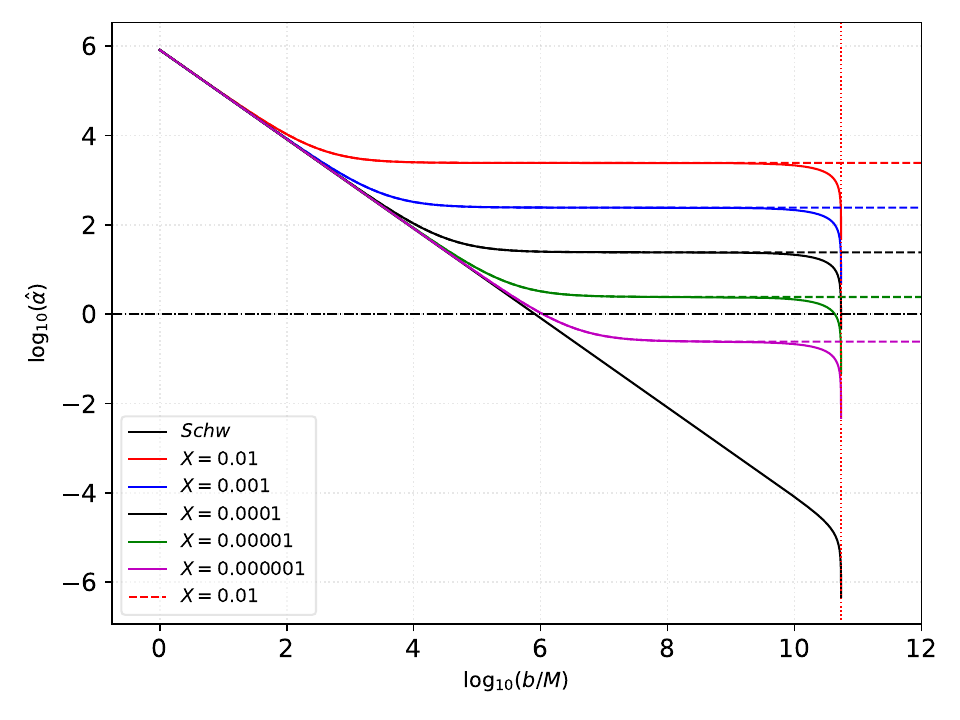}
    \caption{Weak deflection angle (in $\mu$as) using M87* parameters. The left figure compares the behavior of the deflection angle between a massive particle with speed $v = 0.75$(broken lines) and photons where $v=1$ (solid lines). The right plot compares the effect of finite distance to the deflection angle of photons, where the solid lines represent Eq. \eqref{wda1}, and the broken lines represent Eq. \eqref{wda3}. These are all for different theoretical values for the bumblebee parameter $X$. The vertical dotted line represents our location from M87* SMBH, which is $16.8$ Mpc.}
    \label{fig_wda}
\end{figure*}

We saw that the weak deflection angle is sensitive to the metric-affine bumblebee parameter $X$, in contrast to the shadow, which cannot detect the effects of the bumblebee parameter in the strong field regime. To visualize the derived equation, we plot it numerically, and the results are shown in Fig. \ref{fig_wda}. Here, we used some of the M87* SMBH parameters as an example, such as its mass $M = 6.5$x$10^9M_\odot$, and $D = 16.8$ Mpc, which is the distance used between the SMBH and the receiver, as well as the source. It is also useful to plot the deflection behavior in a log-log plot because of these enormous numbers. In the left plot, we observe that time-like particles give a higher value for $\hat{\alpha}$ as the impact parameter of the trajectory lessens. The time-like and null particles give the same $\hat{\alpha}$ value as $b$ becomes comparable to $D$. Next, the bumblebee parameter $X$ increases the value of $\hat{\alpha}$ relative to the Schwarzschild case, but it has a peculiar effect of leveling off this value as $b$ changes. At lower impact parameters, $X$'s effect seems to diminish as it merely follows and coincides with the Schwarzschild behavior. The sensitivity of the weak deflection angle with $X$'s influence is then strong at large values of $b$ leading to its potential detection. Finally, we also plot the comparison between the finite distance effect correction Eq. \eqref{wda1} with the approximated case Eq. \eqref{wda3}. We saw that there was no distinction upon the use of these expressions except when $b$ is nearly the same as $D$.

%%%%%

\section{SHADOWS WITH
INFALLING ACCRETIONS} \label{sec4}

In this section, we use the techniques from references \cite{Jaroszynski:1997bw} and \cite{Bambi:2012tg} to explore a realistic model of the shadow cast by a spherical accretion disk around a black hole. This method accounts for the dynamic nature of accretion disks and their synchrotron emission, which are important factors for obtaining an accurate image of the shadow. We begin by considering the specific intensity of light observed at frequency $\nu_\text{obs}$. This is achieved by solving the integral along the path of the light ray:
        \begin{equation}
            I(\nu_\text{obs},b_\gamma) = \int_\gamma g^3 j(\nu_e) dl_\text{prop}.
            \label{eq:bambiI}
        \end{equation}

The redshift factor accounts for the fact that photons emitted from a free-falling accretion disk are redshifted due to the strong gravitational field of the black hole. The amount of redshift depends on the impact parameter, which is the distance between the photon's trajectory and the center of the black hole. The redshift factor is important for calculating the observed spectrum of an accreting black hole. By taking the redshift factor into account, astronomers can more accurately model the accretion process and learn more about the properties of the black hole. The redshift factor for accretion in free fall is defined as:
        \begin{equation}
            g = \frac{k_\mu u^\mu_o}{k_\mu u^\mu_e}.
        \end{equation}

where where $b_{\gamma}$ is the impact parameter, $j(\nu_e)$ is the emissivity per unit volume, $dl_\text{prop}$ is the infinitesimal proper length, and $\nu_e$  is the frequency of the emitted photon.

In this context, the 4-velocity of the photon is denoted as $k^\mu$, which corresponds to $\dot{x}_\mu$, while the 4-velocity of the distant observer is represented by $u^\mu_o$ and can be written as $(1,0,0,0)$. Additionally, $u^\mu_e$ represents the 4-velocity of the accretion in free fall
\begin{equation}
u_{\mathrm{e}}^{t}=\frac{1}{A(r)}, \quad u_{\mathrm{e}}^{r}=-\sqrt{\frac{1-A(r)}{A(r) B(r)}}, \quad u_{\mathrm{e}}^{\theta}=u_{\mathrm{e}}^{\phi}=0.
\end{equation}

By employing the relation $k_{\alpha} k^{\alpha} = 0$, we can derive constants of motion for photons, namely $k_{r}$ and $k_{t}$.
\begin{equation}
k_{r}=\pm k_{t} \sqrt{B(r)\left(\frac{1}{A(r)}-\frac{b^{2}}{r^{2}}\right)}.
\end{equation}

The redshift factor $g$ tells us how much the photon's frequency is redshifted or blueshifted due to the gravitational field of the black hole.

 \begin{equation}
   g = \Big( u_e^t + \frac{k_r}{k_t}u_e^r \Big)^{-1},
  \end{equation}

  The proper distance $dl_\gamma$ is the distance traveled by the photon along its trajectory, taking into account the curvature of spacetime. When a photon approaches the black hole $+$, it is redshifted. This is because the photon has to lose energy in order to overcome the gravitational pull of the black hole. When a photon moves away $-$ from the black hole, it is blueshifted. This is because the photon gains energy as it escapes the gravitational pull of the black hole.
  
 \begin{equation}
  dl_\gamma = k_\mu u^\mu_e d\lambda = \frac{k^t}{g |k_r|}dr.
\end{equation}

To focus exclusively on monochromatic emission, we can use the specific emissivity with a rest-frame frequency $\nu_*$:

        \begin{equation}
            j(\nu_e) \propto \frac{\delta(\nu_e - \nu_*)}{r^2}.
        \end{equation}

The intensity equation presented in \eqref{eq:bambiI} transforms into the following form for monochromatic emission:
        
\begin{equation}
    F(b_\gamma) \propto \int_\gamma \frac{g^3}{r^2} \frac{k_e^t}{k_e^r} dr.
\end{equation}
        
\begin{figure}[htp]
   \centering
   \includegraphics[scale=0.4]{
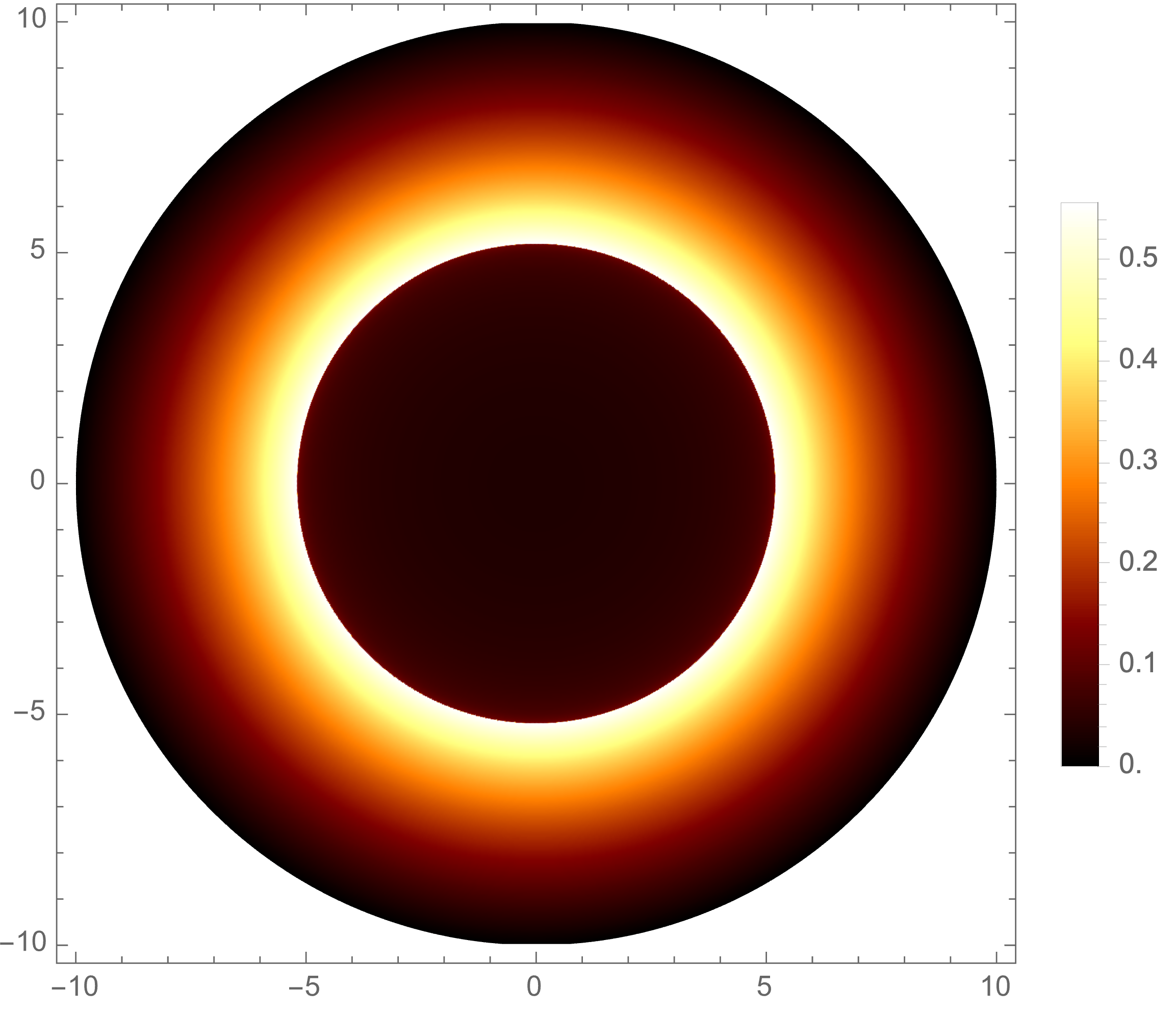}   
\includegraphics[scale=0.4]{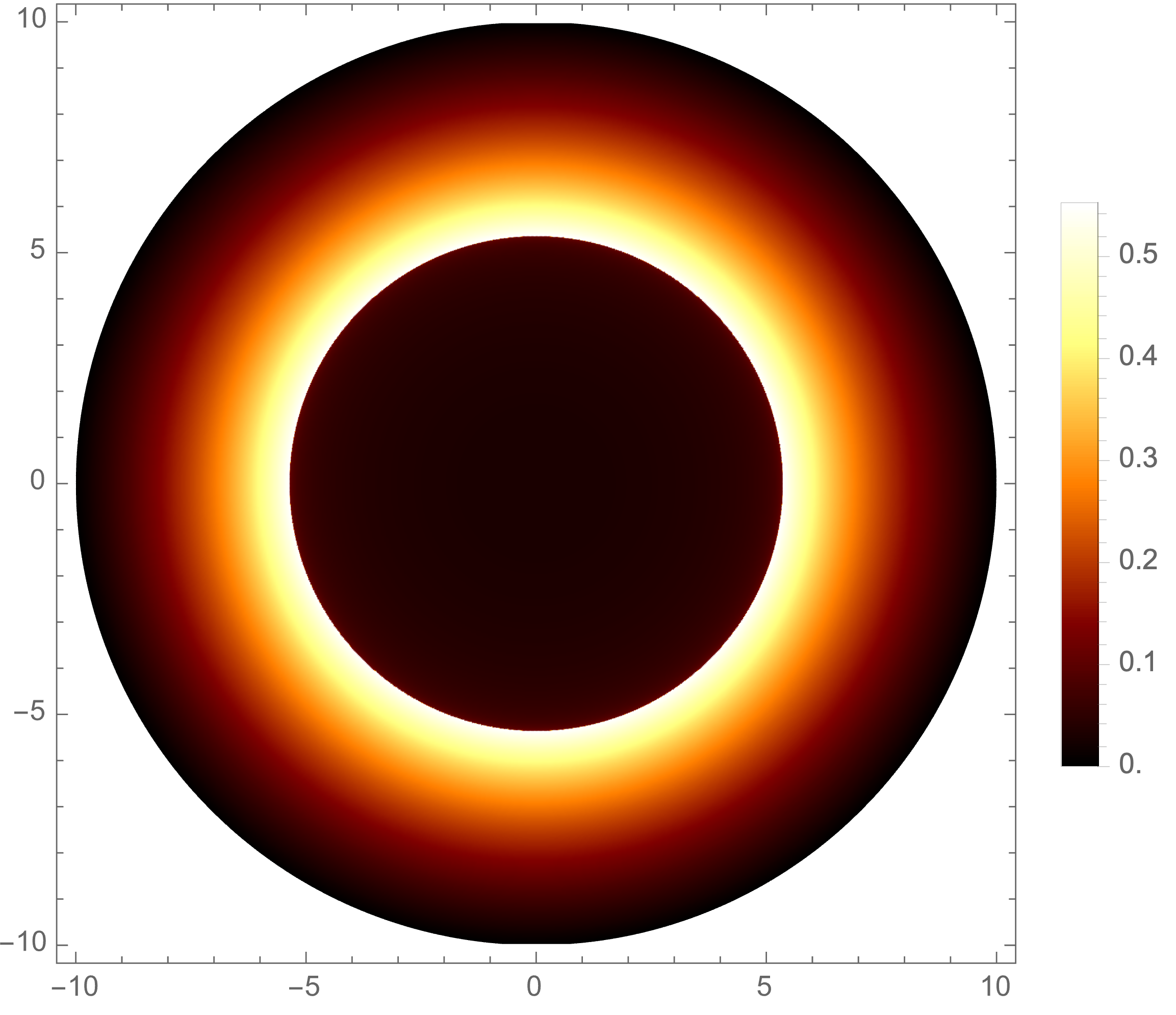}
\includegraphics[scale=0.4]{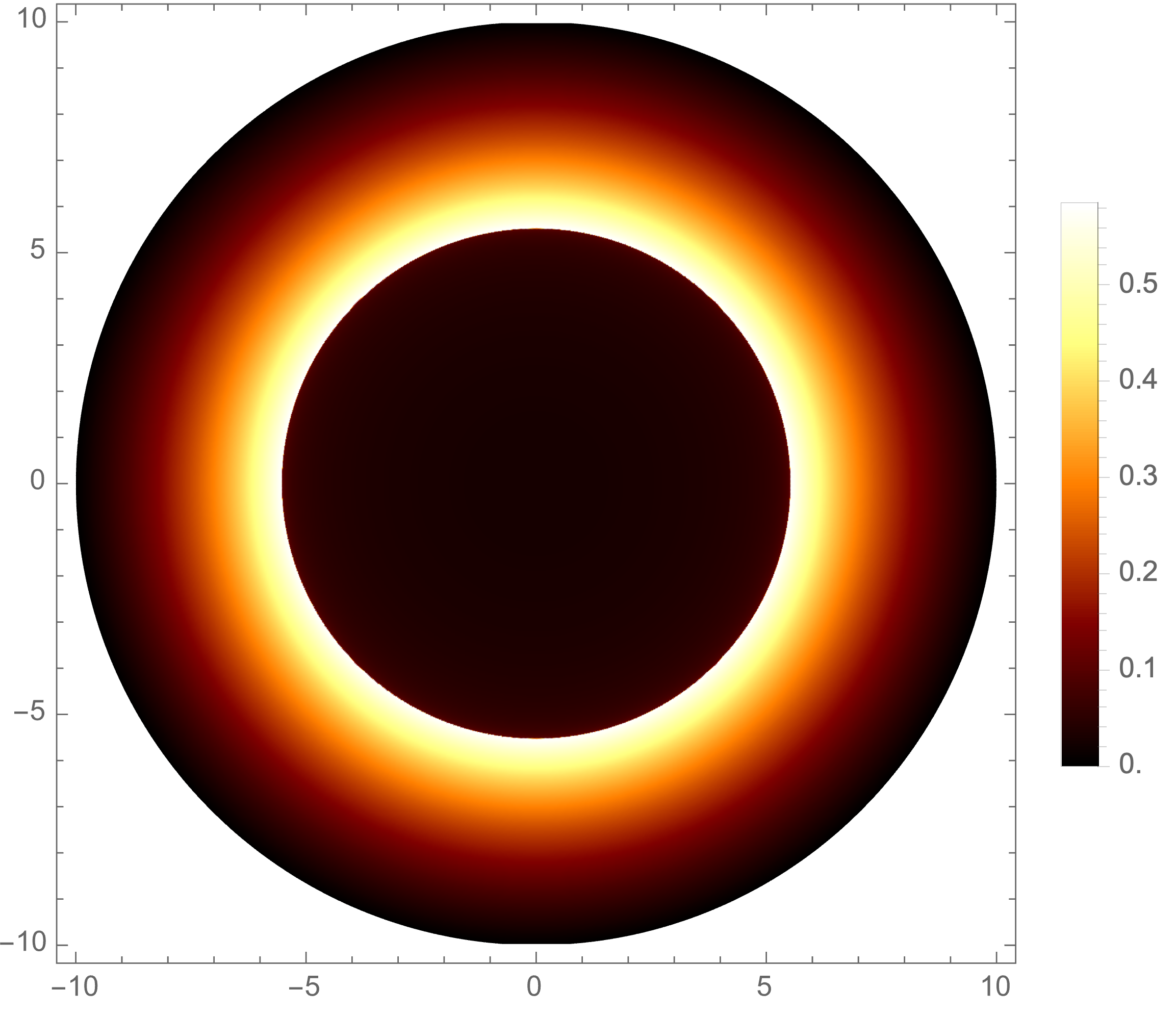}
   \caption{Observational appearance of a spherically free-falling accretion emission near a black hole of charge  $M=1$,variable $X=0.3$,and $X=0.8$ and first one is for Schwarzschild black hole. It is observed that as the value of $X$ increases, the intensity of the emission also increases.}
    \label{fig:shadow.11}
\end{figure}

\begin{figure}[htbp]
    \centering
\includegraphics[width=0.6\textwidth]{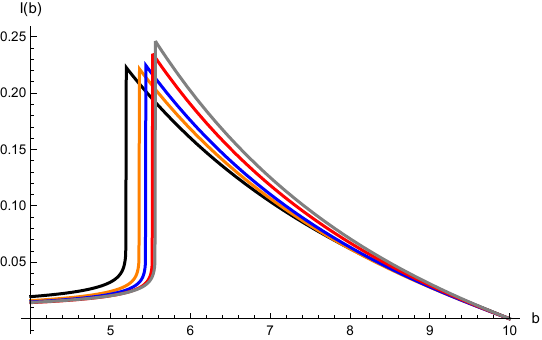}
    \caption{The specific intensity $I_\text{obs}$ seen by a distant observer for an infalling accretion from black hole at fixed $M=1$, and variable $X=0.3$ (orange), $X=0.5$ (blue), $X=0.8$ (red), $X=1$ (gray), and Schwarzschild black hole (black). }
    \label{fig:IntAll}
\end{figure}

We delve into the shadow produced by the thin-accretion disk of Schwarzschild-like black hole in metric-affine Bumblebee gravity framework. To start, we numerically solve the above equation using the \textit{Mathematica} notebook package \cite{Okyay:2021nnh}, which has also been utilized in previous works \cite{Chakhchi:2022fls,Kuang:2022xjp,Uniyal:2022vdu,Pantig:2022gih,Pantig:2022ely,Kumaran:2023brp,Lambiase:2023hng,Uniyal:2023inx,Kumaran:2022soh,Pantig:2022sjb}. This integration of the flux demonstrates how the metric-affine Bumblebee gravity parameter $X$ affects the specific intensity observed by a distant observer for an infalling accretion, as illustrated in Figs. (\ref{fig:shadow.11}, and \ref{fig:IntAll}).

These plots in Figs. (\ref{fig:shadow.11}, and \ref{fig:IntAll}) depict specific intensities for various $X$ values versus the impact parameter $b$ as observed by a distant observer. We observe that increasing the value of $X$ leads to a rise in intensity. Subsequently, the intensity sharply peaks when photons are swiftly captured by the black hole (at the photon sphere). Beyond this peak, intensity gradually diminishes.

The plots (\ref{fig:shadow.11}, and \ref{fig:IntAll}) also show that the intensity peaks when photons are swiftly captured by the black hole (at the photon sphere). This is because the photon sphere is the closest distance to the black hole at which a photon can orbit without falling in. Photons that orbit at the photon sphere are very tightly bound, and therefore emit a lot of light.

Beyond the peak, the intensity gradually diminishes. This is because photons that are farther away from the black hole are less tightly bound, and therefore emit less light.

The plots also show that the intensity is higher for smaller impact parameters. This is because photons with smaller impact parameters have to travel a shorter distance to escape the black hole, and therefore lose less energy. As a result, they are more likely to be observed.

The plots in Figs. (\ref{fig:shadow.11}, and \ref{fig:IntAll}) are important because they provide us with a better understanding of the light emitted by accreting black holes. This information can be used to study the physics of accretion and to constrain the parameters of black holes.

Note that, we indicate that the metric-affine bumblebee gravity parameters cannot be constrained through the black hole shadow. That is, we get the final result as $    R_\text{sh} = 3\sqrt{3}M$.

\section{Greybody factors} \label{sec6} %AO
%\AO{I AM HERE}
The greybody factor (GF) is a parameter that characterizes the probability of a quantum field escaping from a black hole. It is defined as the ratio of the outgoing flux of Hawking radiation to the total flux of Hawking radiation. A high GF indicates a greater likelihood that Hawking radiation can reach infinity.

The GF is important for estimating the intensity of Hawking radiation, which can be used to learn more about the properties of black holes. For example, the GF can be used to constrain the mass and spin of a black hole.  The idea of the rigorous Greybody bound was originally introduced in \cite{Visser:1998ke,Boonserm:2008zg}, offering a qualitative characterization of a black hole. In this section, we will investigate the greybody factor associated with the Schwarzschild-like black hole within the framework of metric-affine Bumblebee gravity.

The greybody factor for a massless scalar field propagating around a Schwarzschild-like black hole in metric-affine Bumblebee gravity can be calculated using the Klein-Gordon equation, which is given by:
\begin{equation}\label{scalar_KG}
\square \Phi = \dfrac{1}{\sqrt{-g}} \partial_\mu (\sqrt{-g} g^{\mu\nu} \partial_\nu \Phi) = 0.
\end{equation}

By disregarding the impact of the field on the spacetime (back-reaction), we can focus solely on Eq. \eqref{pertmetric} at the zeroth order:
\begin{equation}
    ds^2 = -|g_{tt}| dt^2 + g_{rr}dr^2 + r^2 d\Omega_2^2
\end{equation}

The scalar field can be traditionally decomposed using spherical harmonics as follows:
\begin{equation}
\Phi(t,r,\theta, \phi) = \dfrac{1}{r} \sum_{l,m} \psi_l(t,r) Y_{lm}(\theta, \phi),
\end{equation}

Here, $\psi_l(t,r)$ represents the time-dependent radial wave function, with $l$ and $m$ serving as indices for the spherical harmonics $Y_{lm}$. 

The spherical harmonics are a complete set of basis functions for representing scalar fields on a sphere. They are also eigenfunctions of the angular momentum operators.

The radial functions $\psi_l(t,r)$ are determined by solving the Klein-Gordon equation in the radial direction. The solution depends on the angular momentum quantum numbers $l$ and $m$, as well as the mass of the scalar field m and the effective potential $V(r)$. Once the radial functions have been calculated, the scalar field can be reconstructed using the equation above. The decomposition of the scalar field using spherical harmonics is useful for studying the behavior of scalar fields around black holes. For example, the greybody factor for a massless scalar field propagating around a Schwarzschild-like black hole can be calculated using the spherical harmonic decomposition.

The spherical harmonic decomposition is also useful for studying other types of physical systems, such as atoms and molecules.

Substituting these into Eq. \eqref{scalar_KG}, we obtain the following expression:
\begin{equation}
\partial^2_{r_*} \psi(r_*)_l + \omega^2 \psi(r_*)_l = V(r) \psi(r_*)_l,  
\end{equation}

Here, we introduce the tortoise coordinate $r_*$, which is defined as:
\begin{equation}\label{tortoise}
\dfrac{dr_*}{dr} = \sqrt{g_{rr}\, |g_{tt}^{-1}|}
\end{equation}

furthermore, the effective potential of the field, denoted as $V(r)$, is defined as follows:
\begin{equation}\label{Vs}
V(r) = |g_{tt}| \left( \dfrac{l(l+1)}{r^2} +\dfrac{1}{r \sqrt{|g_{tt}| g_{rr}}} \dfrac{d}{dr}\sqrt{|g_{tt}| g_{rr}^{-1}} \right).
\end{equation}

\begin{figure}[htbp]
    \centering
\includegraphics[width=0.6\textwidth]{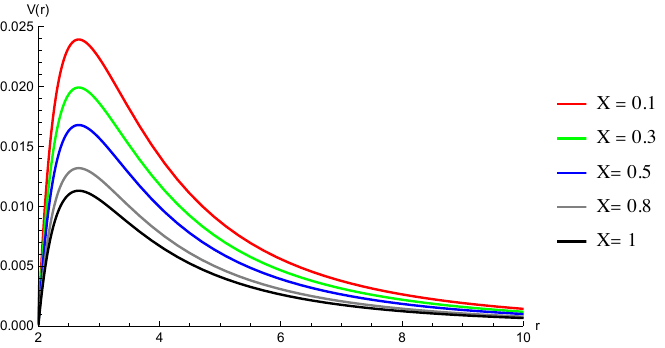}
\includegraphics[width=0.6\textwidth]{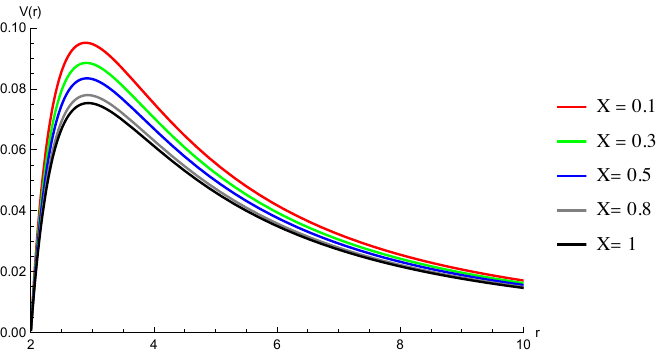}
    \caption{Effective potentials $M=1$, $l=0$ (Top), $l=1$ (Bottom) and variable X. }
    \label{fig:poten}
\end{figure}

The figure \ref{fig:poten} displays the effective potential $V$ as functions of $r$. Notably, the effective potentials exhibit distinct behaviors for variable $X$, and it's evident that they approach the Schwarzschild potentials as $X$ approaches zero. Consequently, we proceed to calculate the bound on the greybody factor

\begin{equation}
T \geq \operatorname{sech}^{2}\left(\int_{-\infty}^{\infty} \vartheta d r_{*}\right),
\end{equation}
where 
\begin{equation}
\vartheta=\frac{\sqrt{\left[h^{\prime}\left(r_{*}\right)\right]^{2}+\left[w^{2}-V\left(r_{*}\right)-h^{2}\left(r_{*}\right)\right]^{2}}}{2 h\left(r_{*}\right)}.
\end{equation}

It's worth noting that the function $h(r_{*})$ fulfills the condition $h(-\infty)=h(\infty)=w,$ as specified in \cite{Visser:1998ke}. By choosing $h=w$ and substituting the tortoise coordinate $r_*$, we can express it as follows:
\begin{equation} \label{bound}
T_{b} \geq \operatorname{sech}^{2}\left(\frac{1}{2 w} \int_{-\infty}^{\infty}\left|V\right| d r  \sqrt{g_{rr}\, |g_{tt}^{-1}|}\right).
\end{equation}

\begin{figure}
    \centering
\includegraphics[width=0.48\textwidth]{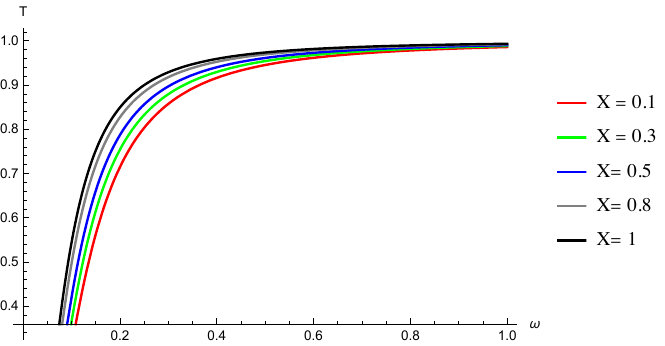}
\includegraphics[width=0.48\textwidth]{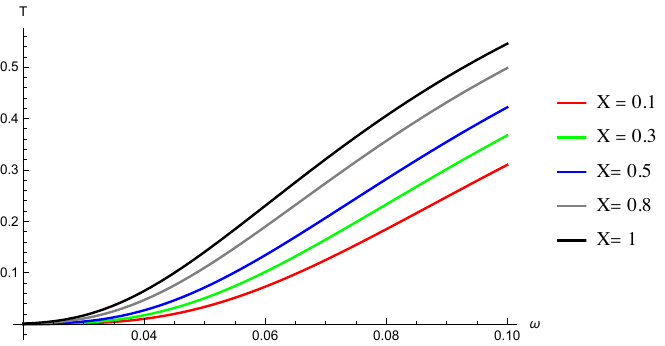}
    \caption{(Scalar Field) The Greybody Bound $T_b$ versus the $\omega$ for different values of values of $X$ parameter, with $M = 1$, $l=0$.}
    \label{fig:greybody1}
\end{figure}

\begin{figure}
    \centering
\includegraphics[width=0.48\textwidth]{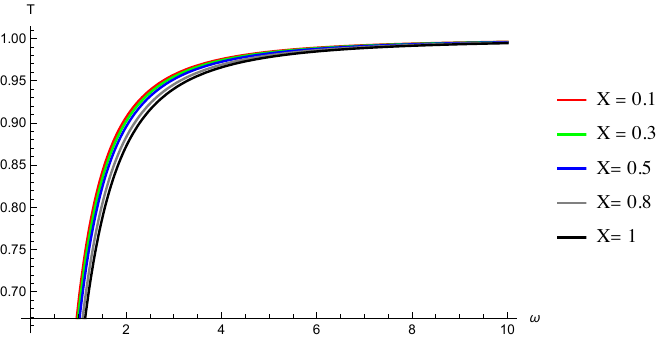}
\includegraphics[width=0.48\textwidth]{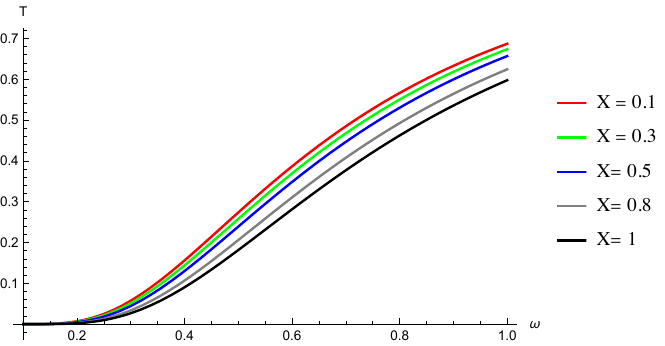}
    \caption{(Scalar Field) The Greybody Bound $T_b$ versus the $\omega$ for different values of values of $X$ parameter, with $M = 1$, $l=1$.}
    \label{fig:greybody2}
\end{figure}

Using the effective potential $V$ for the massless scalar field, we can compute the bound in the following manner:
\begin{equation}
    T \geq T_b = \text{sech}^2\left(\frac{\sqrt{\sqrt{4-X} \sqrt{-\frac{1}{(X-4)^3}}} \left(8 l^2+8 l+\sqrt{\frac{(4-X)^{9/2} \sqrt{-\frac{1}{(X-4)^3}}}{3 X+4}}\right)}{8 M \sqrt{4-X} \omega }\right).
\end{equation}
The bound simplifies to the Schwarzschild case when $(X)\rightarrow 0$, resulting in $T_\text{Sch} \geq \text{sech}^2\left(\frac{2 l (l+1)+1}{8 m w }\right)$. We demonstrate the impact of the screening parameter $X$ on the greybody bound for a scalar field in a Schwarzschild-like black hole within metric-affine Bumblebee gravity in Figures \ref{fig:greybody1} and \ref{fig:greybody2}. Indeed, for $l=0$, it's clear that as the value of the parameter $X$ increases, the greybody bound $T_b$ also increases. Conversely, for $l=1$, as the value of the parameter $X$ increases, the greybody bound $T_b$ decreases.

%%%%

\section{Neutrino energy deposition}
\label{Formulation}

 %The benchmark situation is the final stage of NS merging, which is idealized as BH with an accretion disk. Salmonson \& Wilson in Ref.~\cite{Salmonson:1999es,Salmonson:2001tz} were the first to take into account strong gravitational field effects, in a semi-analytic treatment. They showed that in a Schwarzschild geometry, considering the neutrinos emitted from the central core, the efficiency of the $\nu \bar{\nu}\rightarrow e^{+}e^{-}$ process is enhanced up to a factor of $30$ for collapsing neutron stars (NS) compared to the results in the Newtonian calculation.
%A further study in this direction was considered in \cite{Asano:2000ib,Asano:2000dq}, where the authors studied the general relativistic effects on neutrino pair annihilation near the neutrinosphere and around a thin accretion disk for a Schwarzschild or Kerr metric. Here it was assumed that the accretion disk is isothermal.
%Therefore, let us consider a BH with a thin accretion disk around it that emits neutrinos \cite{Asano:2000dq}. We will confine ourselves to the case of an idealised, semi-analytical, stationary state model, which is independent of details regarding disk formation and where self-gravitational effects are neglected. The disk has an inner and outer edge, with corresponding radii defined by $R_{\mathrm{in}}$ and $R_{\mathrm{out}}$, respectively. The generic metric with a spherical symmetry is given by

The energy deposition rate from the $\nu \bar{\nu}\rightarrow e^{+} e^{-}$ process has been studied to justify the GRB emission. The reference scenario is the final stage of neutron star (NS) merging, conceptualized as a black hole (BH) with an accretion disk. Salmonson and Wilson, as outlined in Ref.~\cite{Salmonson:1999es,Salmonson:2001tz}, were the first to consider the effects of strong gravitational field regimes. They demonstrated that, for a Schwarzschild spacetime and for neutrinos emitted from the central core, the efficiency of the annihilation $\nu \bar{\nu}\rightarrow e^{+}e^{-}$ gets amplified, with respect to the Newtonian counterpart,  by a factor $\sim 30$ for collapsing neutron stars (NS).
In \cite{Asano:2000ib,Asano:2000dq}, the authors investigated the effects of general relativity on neutrino pair annihilation near the neutrinosphere and in the vicinity of a thin accretion disk (assuming an isothermal profile), with the gravitational background described by the Schwarzschild and Kerr geometries.

We consider a black hole (BH) surrounded by a thin accretion disk that emits neutrinos, as discussed in \cite{Asano:2000dq}. We focus on an idealized model which does not depend on the specifics of disk formation and neglects self-gravitational effects. This disk has defined inner and outer edges, corresponding to radii denoted as $R_{\mathrm{in}}$ and $R_{\mathrm{out}}$, respectively. The general metric, exhibiting spherical symmetry, is given by:
\begin{equation}
    g_{\mu\nu}=\left(g_{00},g_{11},-r^2,-r^2\sin^2\theta\right) \,.
\end{equation}
The Hamiltonian can be used to study the motion of the test particle in spacetime. For example, the Hamiltonian allows us to calculate the energy and angular momentum of the test particle, as well as its equations of motion. For a test particle propagating in a curved background, the Hamiltonian is given by
\begin{equation}
2\mathcal{H}=-E\dot{t}+L\dot{\phi}+g_{11}\dot{r}^2=0\,,
\end{equation}
In the given context, where $E$ represents the energy and $L$ signifies the angular momentum of the test particles, the non-vanishing components of the 4-velocity can be derived as follows \cite{Prasanna:2001ie}:
\begin{align}
U^{3}&=\dot{\phi}=-\frac{L}{r^2} \,\ ; \\
U^0&=\dot{t}=-\frac{E}{g_{00}} \,\ ; \\
\dot{r}^2&=\frac{E\dot{t}-L\dot{\phi}}{g_{11}} \,\ ,
\label{dr/dt}
\end{align}
Our focus lies in determining the energy deposition rate in close proximity to the axis, which is perpendicular to the disk, specifically at $\theta=0^{\circ}$. To evaluate the energy emitted within a half cone with an angular extent of approximately $\Delta \theta\sim10^{\circ}$, we need to consider the scalar product of the momenta of a neutrino and an antineutrino at $\theta=0^{\circ}$. This scalar product can be expressed as follows:
\begin{equation}
    p_{\nu}\cdot p_{\bar{\nu}} = E_{\nu}E_{\bar{\nu}}\left[1-\sin\theta_{\nu}\sin\theta_{\bar{\nu}}\cos\left(\phi_{\nu}-\phi_{\bar{\nu}}\right)-\cos\theta_{\nu}\cos\theta_{\bar{\nu}}\right] \,\ ,
\end{equation}
In this context, the term $E_{\nu}$ is defined as the energy of the neutrino, which is given by $E_{0\nu}/\sqrt{g_{00}}$, where $E_{0\nu}$ represents the observed energy of the neutrino at infinity
\begin{equation}
    \sin\theta_{\nu}=\frac{\rho_{\nu}}{r}\sqrt{g_{00}(r)} \,\ ,
    \label{sen}
\end{equation}
Furthermore, it's worth noting that $\rho_{\nu}$ is defined as the ratio of the angular momentum $L_{\nu}$ to the observed energy $E_{0\nu}$. 

Additionally, due to geometric considerations, there are both a minimum and maximum value denoted as $\theta_m$ and $\theta_M$ respectively, for a neutrino originating from $R_{\mathrm{in}}=2R_{\mathrm{ph}}$ and $R_{\mathrm{out}}=30M$, where $R_{\mathrm{ph}}$ is the photosphere radius. In addition, it can be demonstrated that the following relationship holds, as outlined in~\cite{Asano:2000dq}:
\begin{equation}
    \rho_{\nu}=\frac{r_0}{\sqrt{g_{00}(r_0)}} \,\ ,
    \label{rho}
\end{equation}
in which $r_0$ is the nearest position between the particle and the centre before arriving at $\theta=0$. The final component is the trajectory equation, which is presented in~\cite{Asano:2000dq} as follows:
\begin{equation}
    \frac{\pi}{2}=\int_C\frac{dr'}{r'\sqrt{(r'/\rho_{\nu})^2-g_{00}(r')}} \,.
    \label{trajectory equation}
\end{equation}
Equation \eqref{trajectory equation} considers that the neutrinos are emitted from the position $(R,\pi/2)$, where $R$ is within the range $[R_{\mathrm{in}},R_{\mathrm{out}}]$, and then they arrive at $(r,0)$. Consequently, the energy deposition rate resulting from neutrino pair annihilation is given by~\cite{Asano:2000dq}:
\begin{equation}
    \frac{dE_0(r)}{dtdV}=\frac{21\pi^4}{4}\zeta(5)KG_F^2k^9T^9_{\mathrm{eff}}(2R_{ph})F(r) \,\ ,
    \label{trajectory}
\end{equation}
In the given expression, $G_F$ represents the Fermi constant, $k$ stands for the Boltzmann constant, and $T_{\mathrm{eff}}(2R_{\mathrm{ph}})$ denotes the effective temperature at a radius of $2R_{\mathrm{ph}}$ (the temperature observed in the comoving frame)
\begin{equation}
    K=\frac{1\pm 4\sin^2\omega_W+8\sin^4\theta_W}{6\pi} \,\ ,
\end{equation}
in this context, for $\nu_e$, the positive sign is used, while for $\nu_{\mu/\tau}$, the negative sign is applied. Additionally, $\sin^2\theta_W$ represents the Weinberg angle and is equal to 0.23
\begin{equation}
\begin{split}
    &F(r)=\frac{2\pi^2}{T^9_{\mathrm{eff}}(2R_{ph})}\frac{1}{g_{00}(r)^4}\Bigg(2\int_{\theta_m}^{\theta_M}d\theta_{\nu}T_0^5(\theta_{\nu})\sin\theta_{\nu}\int_{\theta_m}^{\theta_M}d\theta_{\bar{\nu}}T_0^4(\theta_{\bar{\nu}})\sin\theta_{\bar{\nu}}+\\&
    +\int_{\theta_m}^{\theta_M}d\theta_{\nu}T_0^5(\theta_{\nu})\sin^3\theta_{\nu}\int_{\theta_m}^{\theta_M}d\theta_{\bar{\nu}}T_0^4(\theta_{\bar{\nu}})\sin^3\theta_{\bar{\nu}}+\\ 
    &+2\int_{\theta_m}^{\theta_M}d\theta_{\nu}T_0^5(\theta_{\nu})\cos^2\theta_{\nu}\sin\theta_{\nu}\int_{\theta_m}^{\theta_M}d\theta_{\bar{\nu}}T_0^4(\theta_{\bar{\nu}})\cos^2\theta_{\bar{\nu}}\sin\theta_{\bar{\nu}}- \\&
    -4\int_{\theta_m}^{\theta_M}d\theta_{\nu}T_0^5(\theta_{\nu})\cos\theta_{\nu}\sin\theta_{\nu}\int_{\theta_m}^{\theta_M}d\theta_{\bar{\nu}}T_0^4(\theta_{\bar{\nu}})\cos\theta_{\bar{\nu}}\sin\theta_{\bar{\nu}}\Bigg) \,\ ,
    \end{split}
    \label{F(r)}
\end{equation}
The term $T_0$ represents the temperature observed at infinity
\begin{align}
    T_0(R)&=\frac{T_{\mathrm{eff}}(R)}{\gamma}\sqrt{g_{00}(R)} \label{Teff} \,\ , \\
    \gamma&=\frac{1}{\sqrt{1-v^2/c^2}} \,\ , \\
    \frac{v^2}{c^2}&=\frac{g_{33}}{g_{00}}\frac{g_{00,r}}{2r} \label{v^2} \,.
\end{align}
$T_{\mathrm{eff}}$ is the effective temperature as measured by a local observer. All quantities are evaluated at $\theta=\pi/2$. In the analysis, the effects of the reabsorption of the deposited energy by the black hole are not considered. We therefore focus on a scenario with a simple temperature gradient, as described in~\cite{Asano:2000dq}:
\begin{equation}\label{tdippr}
T_{\mathrm{eff}}(r)\propto \frac{2R_{ph}}{r} \,\ . 
\end{equation}
The assumptions regarding the temperature values and the shape of the gradient model are consistent with recent findings from neutrino-cooled accretion disk models, as exemplified in references such as \cite{2015ApJ...805...37L,Liu:2007bca,2007ApJ...662.1156K}.

Typically, it is anticipated that the effective maximum temperature, denoted as $T_{\mathrm{eff}}$, falls in the order of $\mathcal{O}(10~\mathrm{MeV})$. This order of magnitude is crucial for achieving the observed neutrino disk luminosity. Consequently, the disk luminosity is not expected to differ significantly across various models. Additionally, since we are not conducting numerical simulations, we assume $T_{\mathrm{eff}}\sim\mathcal{O}(10~\mathrm{MeV})$ to facilitate the comparison of the effects of different gravitational models under the same conditions. It's important to emphasize that despite these theoretical assumptions, the precise temperature profile can only be determined through a disk simulation originating from neutron star (NS) merging with an explicitly defined geometry.

\begin{figure}
    \centering
    \includegraphics[scale=1.2]{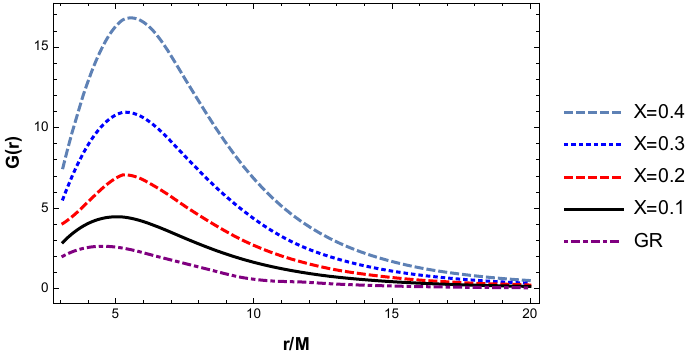}
    \caption{Plot of $G(r)$ vs $r/M$ for different values of $X$ as shown in the legend.}
    \label{fig:EDR}
\end{figure}

In this Section, we have evaluated the energy deposition rate for neutrino annihilation in the affine bumblebee metric described in Eq.~\eqref{eq.metric}. In fig.~\ref{fig:EDR}, we plot the function $ G(r)=F(r) r^2/4M^2$ shown versus the Neutron star radius. It is evident how the parameter $X$ impacts the energy deposition rate at small radius, $r<10~\mathrm{M}$, while $G(r)$ has a similar value with respect to $GR$ at larger radius, $r\sim 20~\mathrm{M}$.\\
The function $G(r)$ plays a pivotal role in computing the energy deposition rate (EDR) and, consequently, in determining the energy available for a GRB (Gamma-Ray Burst) explosion. We calculate the EDR within an infinitesimal angle $d\theta$, considering a characteristic angle of $10^{\circ}$ and a temperature of $10~\mathrm{MeV}$, as outlined in~\cite{Asano:2000dq}:
\begin{equation}
    \frac{dE_0}{dt}\simeq 4.41\times 10^{48}\left(\frac{\Delta \theta}{10^\circ}\right)^2\left(\frac{kT_\mathrm{eff}(R_{\mathrm{in}})}{10~\mathrm{MeV}}\right)^9\left(\frac{2M}{10~\mathrm{km}}\right)\int_{R_{\mathrm{in}}}^{R_{\mathrm{out}}}\frac{G(r)}{2M}dr~\mathrm{erg~s^{-1}} \,\ .
    \label{value}
\end{equation}

Using Eq.~\eqref{value}, one obtains that the energy viable for a GRB emission is
\begin{align}
\frac{dE_0^{GR}}{dt}&\simeq 4.5\times 10^{49}~\mathrm{erg~s^{-1}} \,\ , \\
\mathrm{max}\left(\frac{dE_0^{B}}{dt}\right)&\simeq 3.5\times 10^{50}~\mathrm{erg~s^{-1}} \,\ . 
\label{EDRQuint1}
\end{align}
Many recent time-dependent General Relativity simulations including pair-annihilation and its dynamical impact self-consistently during the evolution~\cite{Fujibayashi_2017,Just:2015dba,PhysRevD.98.063007,2020ApJ...902L..27F} has shown that the process needs at least one order of magnitude for being competitive with the Blandford-Znajek (BZ) process. The usual energy extraction for the BZ process is, indeed, of the order $\sim 6\times 10^{50}~\mathrm{erg/s}$~\cite{Lee:1999se}.
It is hence relevant that the geometrical background provides an enhancement of the energy deposition of almost one order of magnitude with respect to GR in order that the $\nu\bar{\nu}$ annihilation process might be important for the GRB emission. 
%
%We finally remark that, despite all these theoretical hints, only a simulation of the disk from NS-NS merging with assigned geometry would give the exact energy deposition value for the process. 
%
We finally point out that a simulation of the disk from NS-NS merging with assigned geometry is needed in order to give the exact energy deposition value for the processes of GRBs emission. 

\section{CONCLUSIONS}
In this paper, we have investigated Schwarzschild-like black holes within the framework of metric-affine bumblebee gravity. The metric-affine bumblebee gravity theory offers a unique perspective on gravitational interactions by introducing a vector field that couples to spacetime curvature, which can lead to Lorentz symmetry breaking at the Planck scale level. 

In conclusion, our investigation into the effects of metric-affine bumblebee gravity on various astrophysical phenomena has yielded several noteworthy findings.

We have demonstrated that the weak deflection angle is sensitive to the metric-affine bumblebee parameter $X$ while the shadow of the Schwarzschild-like black hole remains unaffected by this parameter in the strong field regime. Through numerical analysis, we have visualized the behavior of the deflection angle in response to different parameters. By using representative values for the M87* supermassive black hole, we observed that time-like particles exhibit higher values of $\hat{\alpha}$ as the impact parameter decreases. The bumblebee parameter $X$ influences $\hat{\alpha}$, particularly at larger impact parameters, highlighting its potential detectability. We have compared the finite distance effect correction equation with the approximated case. In most scenarios, these expressions yield similar results, except when the impact parameter approaches the distance to the observer.

We delved into the specific intensity observed by a distant observer for an infalling accretion and found that increasing the value of $X$ leads to a rise in intensity, peaking at the photon sphere and gradually decreasing thereafter. The impact of the screening parameter $X$ on the greybody bound for a scalar field within a Schwarzschild-like black hole has also been explored. Notably, for different angular momentum quantum numbers ($l=0$ and $l=1$), the behavior of the greybody bound varies with changing $X$ values.

Analyzing the behaviour of accretion disks around Schwarzschild-like black holes in this modified gravity scenario, we explore the disk neutrino energy deposition reaction $\nu\bar{\nu}\rightarrow e^+e^-$. We have computed the energy deposition rate with a link to the GRBs. The release of enormous energy into $e^+ e^-$ and subsequently the annihilation process powers high energetic photons. Using the idealized models of the accretion disk with $T_{\rm eff}\sim r^{-1}$, we have found that the metric-affine bumblebee gravity enhances the energy deposition up to one order of magnitude, with respect to the General Relativity, for values of the free parameter of the theory $X$ such that $X<0.4$. This enhancement is relevant for the gamma-ray burst emission because is of the same order of the BZ process and therefore the neutrino EDR can justify or contribute as a relevant source the observed luminosity of ultra-long GRBs. In summary, our study has provided valuable insights into the intricate interplay between metric-affine bumblebee gravity and astrophysical phenomena, shedding light on the potential detectability of bumblebee parameters and their impact on observables such as deflection angles, intensity profiles, and greybody bounds. These findings contribute to our broader understanding of gravity in the context of modified theories and its implications for black hole astrophysics.

\acknowledgements

The work of G.L. and L.M. is supported by the Italian Istituto Nazionale di Fisica Nucleare (INFN) through the ``QGSKY'' project and by Ministero dell'Istruzione, Universit\`a e Ricerca (MIUR). G.L., A. {\"O}. and R. P. would like to acknowledge networking support by the COST Action CA18108 - Quantum gravity phenomenology in the multi-messenger approach (QG-MM). A. {\"O}. would like to acknowledge networking support by the COST Action CA21106 - COSMIC WISPers in the Dark Universe: Theory, astrophysics and experiments (CosmicWISPers).

\bibliography{references.bib}

\end{document}